\documentstyle[12pt,epsf,epsfig,wrapfig]{article}
\begin{document}
\hfill RUB-TPII-16/05
\begin{center}
{\bfseries IMPROVING QCD PERTURBATIVE PREDICTIONS AND
           TRANSVERSE MOMENTUM EFFECTS%
\footnote{Invited talk at the XI Workshop on High Energy Spin Physics,
          Dubna, Russia, September 27 - October 1, 2005
}}

\vskip 5mm
N.G. Stefanis$^{\dag}$

\vskip 5mm
{\small
{\it
Institut f\"{u}r Theoretische Physik II, Ruhr-Universit\"{a}t Bochum,
D44780 Bochum, Germany
}\\
$\dag$ {\it
E-mail: stefanis@tp2.ruhr-uni-bochum.de
}}
\end{center}

\vskip 3mm
\begin{abstract}
I elaborate on the implementation of analyticity of hadronic quantities
at the amplitude level and show that it amounts to the extension of
Analytic Perturbation Theory of Shirkov and Solovtsov to non-integer
(fractional) powers of the strong running coupling.
I give evidence that at the NLO of the factorized pion's
electromagnetic form factor this framework diminishes the uncertainty
due to the renormalization-scheme choice and nearly removes the
dependence on both the renormalization and the factorization scale.
I supplement this discussion by considering effects caused by retaining
transverse momenta that give rise to Sudakov factors.
\end{abstract}

\vskip 3mm
QCD perturbation theory is an area of research whose results have
proven mandatory to make predictions for various hard hadronic
collisions.
However, the higher the order of the perturbative expansion, the more
complex and difficult it becomes to ensure its self-consistency and
precision.
One reason is that such a procedure may bear through its truncation a
strong dependence on the choice of the renormalization scheme adopted
and the renormalization scale chosen.
Moreover, the fixed-order perturbative calculation is sensitive to the
variation of the factorization scale as well.
Factorization is an indispensable tool to `take out' of the calculation
that fraction of the dynamics that can be described by quark-gluon
exchanges governed by perturbative QCD (pQCD).
The remainder of the reaction amplitude is attributed to nonperturbative
interactions and has to be modelled or extracted from experiment.

This talk begins with a brief summary of the key features
pertaining to the issue of analyticity, using as an example the pion's
electromagnetic form factor.
Then, it explores the ways the perturbative treatment can be improved.
This is achieved by imposing an analyticity requirement on the reaction
amplitude as a \emph{whole}, as postulated by Karanikas and Stefanis
(KS) \cite{KS01}.
This analyticity requirement naturally generalizes the original
Shirkov-Solovtsov approach \cite{SS97}, which replaces the running
coupling and its powers by analytic expressions in which the
Landau pole is absent by construction.
Until now the power-series expansion in $\alpha_{\rm s}(Q^2)$ is the
predominant paradigm for perturbative calculations, one strong argument
in its favor being standardization.
But growing precision of experimental data is putting great pressure on
perturbative calculations, demanding the reduction of
uncertainties ensuing from non-negligible corrections at the subsequent
order in perturbation theory and a slow convergence of the truncated
expansion.
Trading the power-series expansion for a functional expansion in terms
of analytic images of the coupling and its powers---in both the
Euclidean and the Minkowski space---according to the rules of Analytic
Perturbation Theory (APT) \cite{SS97,Shi98}, one can improve the
convergence of the series and expand hadronic quantities in the whole
complex $Q^2$ plane.

Yet, there is a significant limitation of this approach to integer
powers of the coupling.
As it turns out \cite{BKS05}, logarithms, depending on the factorization
(or evolution) scale---typical in higher-order perturbative
calculations---though do not change the Landau pole, affect the
discontinuity across the cut along the negative real axis
$-\infty<Q^2<0$ and have, therefore, to be included into the
``analytization'' procedure, i.e., the spectral density.
Their analytic images correspond to non-integer (fractional) powers
of the running coupling.
The full implementation of this formalism, dubbed Fractional Analytic
Perturbation Theory, or FAPT\ for short, has been just completed in
\cite{BMS05}.
FAPT\ is valuable also in resumming powers of the running coupling
under the proviso of analyticity, allowing to discuss the key
features of Sudakov resummation beyond the naive ``analytization''
approach developed in \cite{SSK98}.

After this brief introduction, let me present the analytic machinery
in more detail.
The analytic conversion proceeds via the general prescription
\begin{eqnarray}
 \left[f(Q^2)\right]_{\rm an}
  =
  \frac{1}{\pi}
   \int_0^{\infty}\!
    \frac{\textbf{Im}\,\big[f(-\sigma)\big]}
         {\sigma+Q^2-i\epsilon}\,
     d\sigma\,,
\label{eq:dispersion}
\end{eqnarray}
%%Eq (1)
which connects at loop order $(l)$ the conventional powers of the
running coupling $a_{(l)}^{m}(Q^2)$, marked $m$, to analytic
expressions ${\cal A}^{(l)}_{m}(Q^2)$ with index $m$.
Imposing analyticity of the pion form factor in the generalized sense
of KS \cite{KS01} leads at NLO to the following expression for the
hard-scattering (perturbative) amplitude:
\begin{eqnarray}
 && \left[Q^2 T_{\rm H}\left(x,y,Q^2;\mu^{2}_{\rm F},\lambda_{\rm R} Q^2\right)
 \right]_{\rm KS}^{\rm an}
  = {\cal A}_{1}^{(2)}\left(\lambda_{\rm R} Q^2\right)\, t_{\rm H}^{(0)}(x,y)
 ~~~~~~~~~~~~~~~~~~~~~~~~~~~~~~~~~~~~
 \nonumber\\ %[0.5cm]
&& ~~~~~~~~~~ +\ \frac{{\cal A}_{2}^{(2)}\left(\lambda_{\rm R} Q^2\right)}{4\pi}\,
       \left[b_0\,t_{\rm H}^{(1,\beta)}(x,y;\lambda_{\rm R})
           + t_{\rm H}^{({\rm FG})}(x,y)
           + C_{\rm F}\,
             t_{{\rm H},2}^{(1,{\rm F})}
             \left(x,y;\frac{\mu^{2}_{\rm F}}{Q^2}\right)
       \right]
       \nonumber\\
&& ~~~~~~~~~~ +\ \frac{\Delta_{2}^{(2)}
  \left(\lambda_{\rm R} Q^2\right)}{4\pi}\,
      \left[C_{\rm F}\, t_{\rm H}^{(0)}(x,y)  \,
             \left(6 + 2 \ln(\bar{x}\bar{y})\right)
      \right]\, ,\!\!\!\!\!\!\!\!\!\!\!\!\!\!\!
\label{eq:TH-KS-6}
\end{eqnarray}
%%Eq (2)
where the deviation from its counterpart within the
Shirkov-Solovtsov scheme, using the \emph{maximal} ``analytization''
procedure \cite{BPSS04}, is encoded in the term \cite{BKS05}
\begin{eqnarray}\label{eq:delta2-2}
 \Delta_{2}^{(2)}\left(Q^2\right)
  &\equiv&
   {\cal L}_{2}^{(2)}\left(Q^2\right)
    - {\cal A}_{2}^{(2)}\left(Q^2\right)\,\ln\left[Q^2/\Lambda^2\right]
\end{eqnarray}
%%Eq (3)
with
\begin{eqnarray}\label{eq:Log_Alpha_2_KS}
 {\cal L}_{2}^{(2)}\left(Q^2\right)
  &\equiv&
   \left[\left(\alpha_{s}^{(2)}\left(Q^2\right)\right)^2
         \ln\left(\frac{Q^2}{\Lambda^2}\right)
   \right]_{\rm KS}^{\rm an}
  =  \frac{4\pi}{b_0}\,
     \left[\frac{\left(\alpha_{s}^{(2)}\left(Q^2\right)\right)^2}
                 {\alpha_{s}^{(1)}\left(Q^2\right)}
      \right]_{\rm KS}^{\rm an}\, .
\end{eqnarray}
%%Eq (5)
Here I have written
$\ln(Q^2/\mu^{2}_{\rm F})
=
 \ln (\lambda_{\rm R} Q^2/\Lambda^2) -
 \ln (\lambda_{\rm R}\mu^{2}_{\rm F}/\Lambda^2)$
with the renormalization scale set to
$\mu_{\rm R}^{2}=\lambda_{\rm R}Q^2$ and $\bar{x}\equiv 1-x$.
The subscript KS in the last displayed equation reminds us that this
expression should be analyticized according to the KS prescription.
This means that the dispersive image of Eq.\ (\ref{eq:Log_Alpha_2_KS})
must include not only the couplings but also the logarithmic terms.
For further details and explicit expressions, I refer to \cite{BKS05}
and \cite{BPSS04}.
The final result of the KS ``analytization'' is
\begin{eqnarray}\label{eq:Log_Alpha_2_BMKS}
  {\cal L}_{2}^{(2)}\left(Q^2\right)
   = \frac{4\pi}{b_0}\,
      \left[{\cal A}_{1}^{(2)}\left(Q^2\right)
      + c_1\,\frac{4\pi}{b_0}\,f_{\cal L}\left(Q^2\right)
      \right]\, ,
\end{eqnarray}
%%Eq BKS (3.18) ---> (4)
where
\begin{eqnarray}\label{eq:f_MS}
  f_{\cal L}\left(Q^2\right)
   = \sum_{n\geq0}
      \left[\psi(2)\zeta(-n-1)-\frac{d\zeta(-n-1)}{dn}\right]\,
       \frac{\left[-\ln\left(Q^2/\Lambda^2\right)
             \right]^n}{\Gamma(n+1)}
\end{eqnarray}
%%Eq (6)
and $\zeta(z)$ is the Riemann zeta-function.
One can calculate the factorizable pion's electromagnetic form factor
by convoluting Eq.\ (\ref{eq:TH-KS-6}) with the pion distribution
amplitude (DA), which gives the probability for finding two collinear
partons with fractions $x_{1}=x$ and $x_{2}=1-x$ of the pion's
momentum:
\begin{equation}
  F_{\pi}^{\rm Fact}(Q^{2};\mu^{2}_{\rm R})
=
  \int_{0}^{1}[dx][dy]
  \Phi_{\pi}^{*}\left(x,\mu_{\rm F}^{2}\right)
  T_{\rm H}\left(x,y,Q^2;\mu^{2}_{\rm F},\lambda_{\rm R} Q^2\right)
  \Phi_{\pi}\left(y,\mu_{\rm F}^{2}\right)\, ,
\label{eq:pi-FF-fact}
\end{equation}
%%Eq (7)
where $[dx]=dx_{1}dx_{2}\delta(x_{1}+x_{2}-1)$ and
$\Phi_{\pi}$ represents the nonperturbative QCD input.
Here, I use a model DA, derived from nonlocal QCD sum rules
\cite{BMS01}, that was found \cite{BMS02} to provide best agreement
with the CLEO and CELLO data on the pion-photon transition
(corresponding references are given in \cite{BMS02}).
In Fig.\ 1, I display the ratio of the pion form factor, calculated
with the KS ``analytization'' prescription, relative to the expression
following from the \emph{maximal} one \cite{BPSS04} (see below), and
using three different scheme- and scale settings.
One sees from this figure that using the default
$\overline{{\rm MS}\vphantom{^1}}$ scheme, the results of both
``analytization'' procedures yield almost coincident results, whereas
in the BLM\ scheme and also in the $\alpha_V$ scheme, the KS prediction
comes out smaller by a few percent.
Obviously, the inherent theoretical uncertainties due to the involved
perturbative parameters, defining a renormalization scheme and scale
setting, are further reduced.
A second important feature of the KS procedure is that the
dependence of $F_{\pi}^{\rm Fact}(Q^2)$ on the factorization
scale is almost diminished, as it is illustrated in Fig.\ 3 of Ref.\
\cite{BKS05}.

%%%%%%%%%%%%%%%%%%%%%%%%%%%%%%%% Fig. 1 %%%%%%%%%%%%%%%%%%%%%%%%%%%%%%%
\begin{figure}[b!]
\begin{center}
\begin{tabular}{ccc}
\mbox{\epsfig{figure=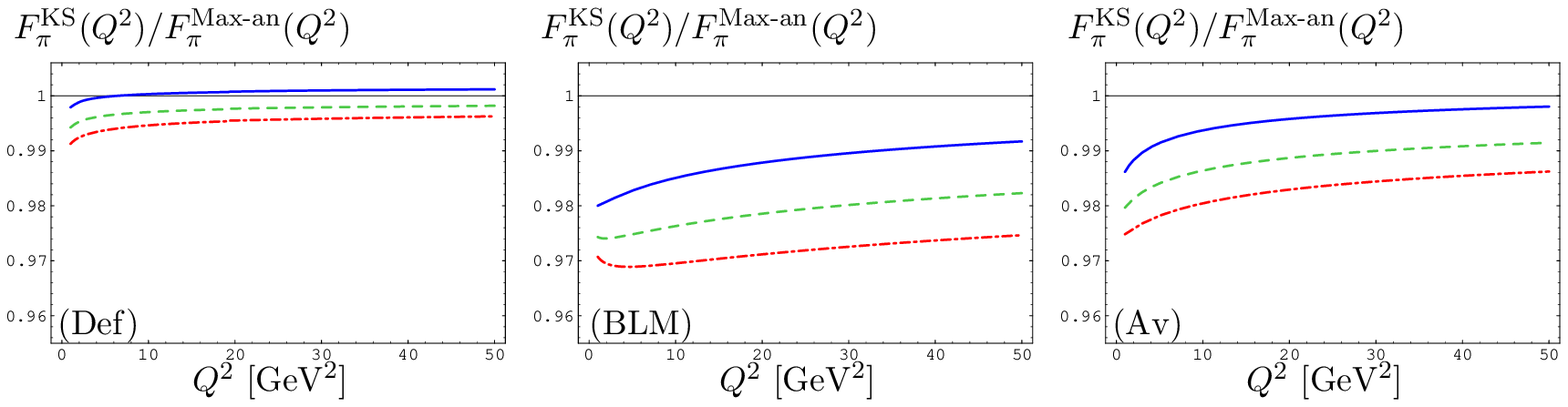,width=15cm,height=4cm}}&
%\mbox{\epsfig{figure=yourname_fig2b.eps,width=4cm,height=2cm}}&
%\mbox{\epsfig{figure=yourname_fig2c.eps,width=4cm,height=2cm}}\\
%{\bf(a)}& {\bf(b)}& {\bf(c)}
\end{tabular}
\end{center}
{\small{\bf Figure 1.} KS ``analytization'' in comparison with the
\emph{maximal} ``analytization'' in terms of the ratio
$F^{\rm KS}_{\pi}(Q^2)/F^{\rm Max-an}_{\pi}\left(Q^2\right)$ for
three different scale settings explained in the text.}
\label{stefanis_fig1}
\end{figure}
%%%%%%%%%%%%%%%%%%%%%%%%%%%%%%%%%%%%%%%%%%%%%%%%%%%%%%%%%%%%%%%%%%%%%%%

Up to now, I have discarded transverse degrees of freedom.
Retaining the $k_\perp$-dependence of the hard-scattering amplitude and
reverting to the pion wave function before integrating over $k_\perp$,
one has a modified factorization that explicitly takes into account
Sudakov factors for each incoming and outgoing parton involved in the
process. %\cite{LS92}.
The first attempt to extend the ``analytization'' principle of Shirkov
and Solovtsov to the Sudakov summation of logarithms in QCD, was
undertaken in \cite{SSK98} and refined in \cite{SSK00}.
The Sudakov factor is defined in terms of the momentum-dependent cusp
anomalous dimension, viz.,
\begin{equation}
  s\left(\xi , b, Q, C_{1}, C_{2} \right)
=
  \frac{1}{2}
  \int_{C_{1}/b}^{C_{2}\xi Q}
  \frac{d\mu}{\mu} \,
  \Gamma _{{\rm cusp}}
        \left(\gamma , \alpha _{\rm s}^{\rm an}(\mu )
        \right)\, ,
\label{eq:sudfuncusp}
\end{equation}
%%Eq (8)
where
\begin{eqnarray}
  \Gamma _{{\rm cusp}}
        \left( \gamma , \alpha _{\rm s}^{\rm an}(\mu ) \right)
& = &
    \Gamma _{{\rm cusp}}^{\rm pert}
  +
    \Gamma _{{\rm cusp}}^{\rm npert}
\label{eq:gammacusp}
\end{eqnarray}
%%Eq (9)
with
$\gamma = \ln \left(C_{2}\xi Q/\mu \right)$,
and each term pertains to the corresponding contribution in the
analyticized running coupling at the two-loop level.
The Sudakov factor resums logarithms with gluon virtualities confined
in the range limited by $C_{1}/b$ (transversal or IR cutoff) and
$C_{2}\xi Q$ (collinear or longitudinal cutoff), where
$\xi = x, \bar{x}, y, \bar{y}$ and $b$ is the impact parameter (i.e.,
the transverse distance between the two partons of the pion).
%%%%%%%%%%%%%%%%%%%%%%%%%%%%%%%% Fig. 2 %%%%%%%%%%%%%%%%%%%%%%%%%%%%%%%
\begin{wrapfigure}[20]{t,R}{5cm}
\begin{center}
\vspace{-5mm}
\mbox{\epsfig{figure=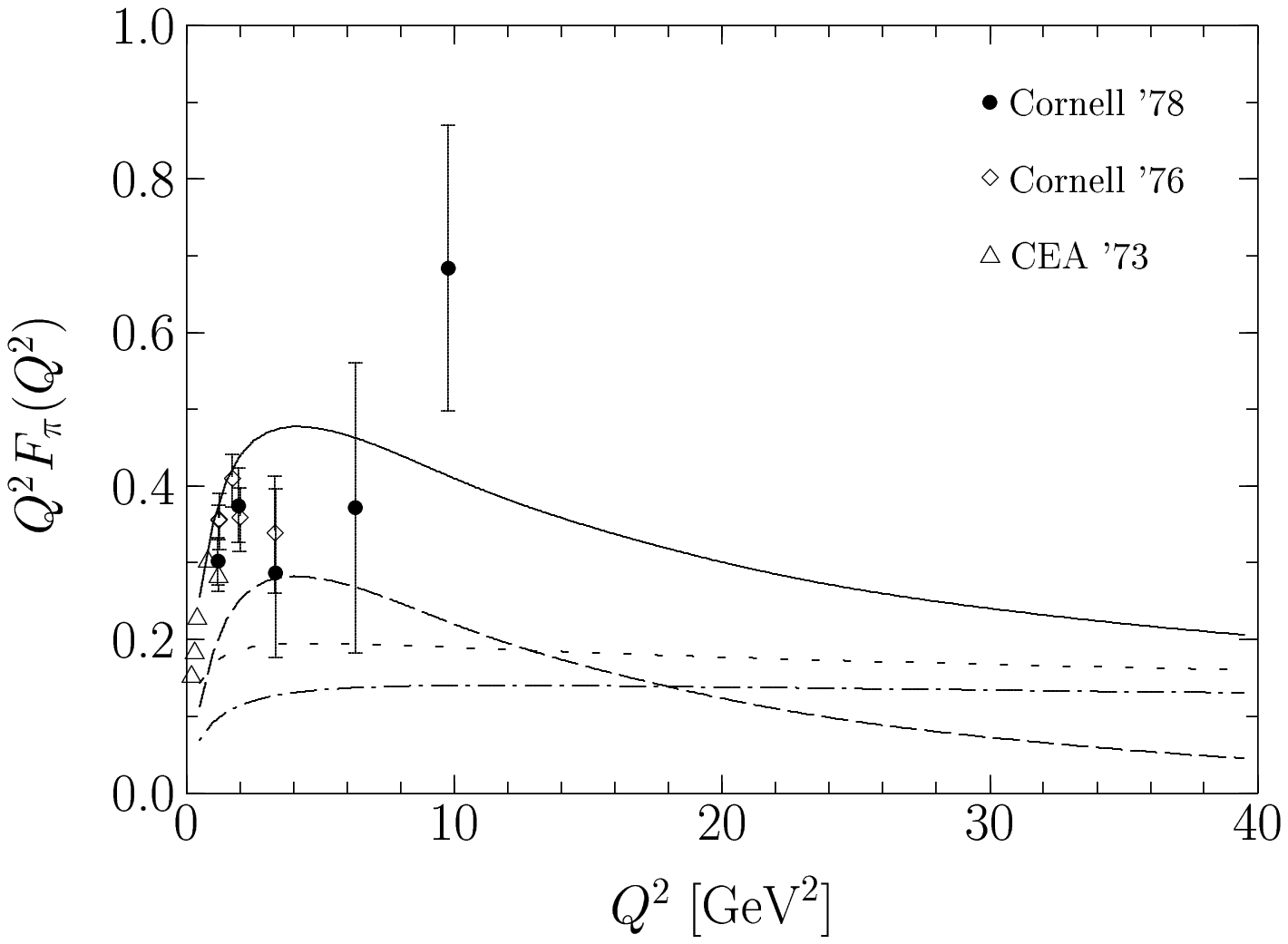,width=5cm,height=4cm}}
\end{center}
\vspace{-1mm}
{\small{\bf Figure 2.} Pion form factor incorporating Sudakov
   effects with \emph{naive} ``analytization'' \protect\cite{SSK00}.
   Dotted line--LO result; dash-dotted line--NLO result; dashed
   line--soft contribution; solid line--sum of NLO contribution and
   soft one.
   }
\label{stefanis_fig2}
\end{wrapfigure}
%%%%%%%%%%%%%%%%%%%%%%%%%%%%%%%%%%%%%%%%%%%%%%%%%%%%%%%%%%%%%%%%%%%%%%%

The scheme constants $C_{i}$ emerge from the truncation of the
perturbative series and would be absent in all-order expressions in
the coupling constant. (For a detailed discussion of these constants
and the explicit expressions for the two Sudakov contributions, I
refer to \cite{SSK00}).
The Sudakov factor receives now additional sub-leading logarithmic
contributions originating from the pole remover in the analyticized
running coupling.
The main effect of these contributions is to provide IR enhancement
to the pion form factor already in the range of currently probed
momentum-transfer values.
This is because the range in which soft gluons build up the Sudakov
factor is enlarged and inhibition of bremsstrahlung sets in at larger
$Q^{2}$ values.
The prediction within this theoretical framework, taken from
\cite{SSK00}, is shown in Fig.\ 2.

Before finishing this section, I discuss an ambiguity which affects
the presented analysis.
Above, we have replaced
$(\alpha _{\rm s}^{(1)})^{2}$ by
$(\alpha _{\rm s}^{{\rm an}(1)})^{2}$ (\emph{naive} ``analytization''
\cite{SSK98,SSK00}),
but exact ``analytization'' would require to analyticize instead
the square of the running coupling as in the \emph{maximal}
``analytization'' procedure \cite{BPSS04}:
$
 (\alpha _{\rm s}^{(1)})^{2}|_{\rm an}
$.
Hence, the question arises: What means (Sudakov) resummation
of powers of the running coupling while maintaining analyticity?
Because the analytic images of these powers translate into a
non-power-series expansion, the result of this operation is no
more simple exponentiation.
This interesting finding was obtained recently in \cite{BMS05}
within the framework of FAPT.
For illustration, I shall only display here graphics for a toy model
of the Sudakov factor that, nevertheless, bears the salient
characteristics of the true one.

Figure 3 shows the expression
$F_{\rm Sud}(x,L)\equiv \exp{\left[-x\,a^{(l)}(L)\right]}$,
$x$ being a free parameter,
$a^{(l)}=\frac{b_0\alpha_{s}^{(l)}}{(4\pi)}$, and
restricting the loop order to $l=1$.
Then, for \hbox{$L=\ln(Q^2/\Lambda^2)>0$}, one obtains
\begin{eqnarray}
\label{eq: exp-a1}
  F_{\rm S}^{(\rm an)}(x, L)
=
    e^{-x/L}
  + \sqrt{x}\sum_{m=1}e^{-L\, m}\,\frac{J_1(2\sqrt{x m})}{\sqrt{m}}
\, .
\end{eqnarray}
%%Eq (10)
Above, the perturbative part of the analytic one-loop running
coupling reproduces exactly the asymptotic expression
$\exp\left[- x\, a^{(1)}(L)\right]$, while the Landau pole
remover generates the sum of the exponents in $-L$, weighted by
the Bessel functions, $J_1$, exhibiting how the large $L$
behavior is violated.\\
%%%%%%%%%%%%%%%%%%%%%%%%%%%%%%%% Fig. 3 %%%%%%%%%%%%%%%%%%%%%%%%%%%%%%%
\begin{wrapfigure}[16]{T,R}{5cm}
\vspace{-9mm}
\begin{center}
\mbox{\epsfig{figure=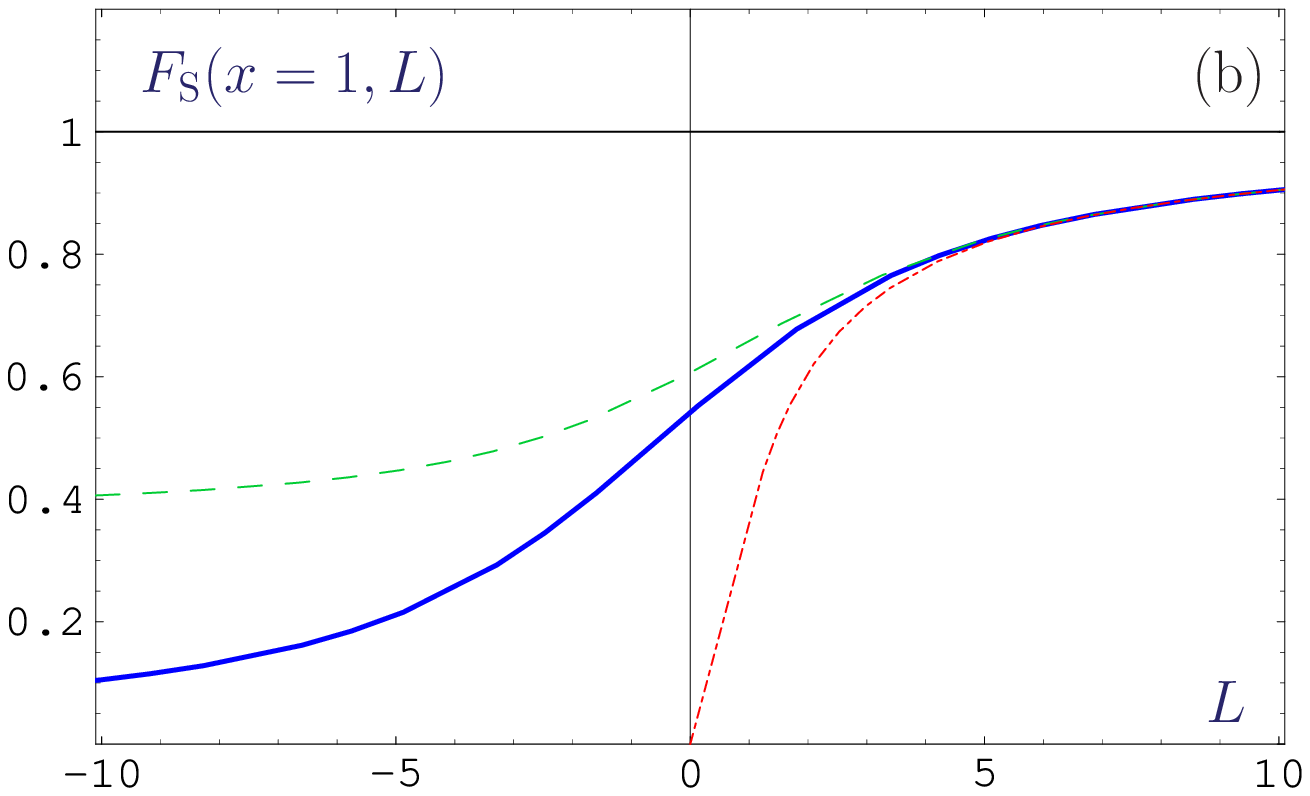,width=5cm,height=4cm}}
\end{center}
\vspace{-3mm}
{\small{\bf Figure 3.} Toy Sudakov function within
FAPT
    (solid line), result of the \emph{naive} ``analytization''
    \cite{SSK00} (dashed line), and standard pQCD (dotted line).
    Figure taken from \protect\cite{BMS05}.
    }
\label{stefanis_fig3}
\end{wrapfigure}
%%%%%%%%%%%%%%%%%%%%%%%%%%%%%%%%%%%%%%%%%%%%%%%%%%%%%%%%%%%%%%%%%%%%%%%

\vspace{-5mm}
In conclusion, I have presented an analytic framework for calculating
hadronic amplitudes, in which not only the running coupling and its
powers, but also logarithms are integral parts of the ``analytization''
procedure.
This so-called FAPT naturally generalizes the original APT of
Shirkov-Solovtsov to fractional powers of the coupling and allows
one to adopt any renormalization scheme and scale setting without
introducing uncertainties, while the calculation becomes also
insensitive to the variation of the factorization scale.

I would like to thank A.P. Bakulev, A.I. Karanikas, S.V. Mikhailov,
and W. Schroers for collaboration on the issues presented in this talk.
I am also grateful to the Deutsche Forschungsgemeinschaft for a travel
grant.
This work was supported in part by the Heisenberg-Landau Programme.

\end{document}